\documentclass{article}

\usepackage{hyperref}

\usepackage{amsmath}
\usepackage{amssymb}
\usepackage{amsthm}
\usepackage[T1]{fontenc}
\theoremstyle{definition}
\newtheorem{definition}{Definition}
\newtheorem{example}{Example}
\newtheorem{theorem}{Theorem}

\newtheorem{corollary}{Corollary}

\usepackage{graphicx}
\usepackage{tikz}
\usetikzlibrary{arrows.meta,positioning,shapes.misc}
\usetikzlibrary{decorations.pathreplacing,patterns,positioning}
\usetikzlibrary{arrows}
\usetikzlibrary{calc}
\usepackage{xcolor}
\usepackage{ulem}

\providecommand{\keywords}[1]
{
  \small	
  \textbf{\textit{Keywords---}} #1
}

\title{Weighted majority tournaments and Kemeny ranking with 2-dimensional Euclidean preferences}%Kemeny ranking is NP-hard for 2-dimensional Euclidean preferences}
\date{}

\author{Bruno Escoffier$^{1,2}$, Olivier Spanjaard$^1$, Magdal\'ena Tydrichov\'a$^1$\\
\small $^1$ Sorbonne Universit\'e, CNRS, LIP6, 4 place Jussieu, 75005 Paris, France\\
\small $^2$ Institut Universitaire de France (IUF), France\\
\small \texttt{name.surname@lip6.fr}
}

\begin{document}
\maketitle

\begin{abstract}
    The assumption that voters' preferences share some common structure is a standard way to circumvent NP-hardness results in social choice problems. While the Kemeny ranking problem is NP-hard in the general case, it is known to become easy if the preferences are 1-dimensional Euclidean. In this note, we prove that the Kemeny ranking problem remains NP-hard for $k$-dimensional Euclidean preferences with $k\!\ge\!2$ under norms $\ell_1$, $\ell_2$ and $\ell_\infty$, by showing that any weighted tournament (resp. weighted bipartite tournament) with weights of same parity (resp. even weights) is inducible as the weighted majority tournament of a profile of 2-Euclidean preferences under norm $\ell_2$ (resp. $\ell_1,\ell_{\infty}$), computable in polynomial time. More generally, this result regarding weighted tournaments implies, essentially, that hardness results relying on the (weighted) majority tournament that hold in the general case (e.g., NP-hardness of Slater ranking) are still true for 2-dimensional Euclidean preferences.
\end{abstract}


%\linenumbers

\section{Introduction}
\normalem

Aggregation rules are ubiquitous in social choice theory \cite{arrow2012social}. Given a multiset of rankings of candidates, an aggregation rule returns a \emph{consensus ranking}, i.e., a ranking that fairly reflects the various preferences expressed in the input rankings. One of the most popular aggregation rules is the Kemeny rule, that returns a ranking minimizing the sum of Kendall tau distances to the input rankings \cite{kemeny1959mathematics} (the Kendall tau distance between two rankings is the number of pairwise disagreements between them). 

While the computation of a consensus ranking can be performed in polynomial time in the number of votes and candidates for some voting rules (e.g., the Borda rule), it is well-known that determining a consensus ranking for the Kemeny rule (problem named \emph{Kemeny ranking} in the following) is NP-hard in the general case \cite{bartholdi1989voting}. A standard way to circumvent this drawback is to assume that the preferences are \emph{structured} \cite{elkind2017structured}. Structured preferences include, for instance, single-peaked preferences \cite{Black58}, single-crossing preferences \cite{roberts1977voting},
1-Euclidean preferences \cite{hotelling1929stability,coombs1950}, or $k$-Euclidean preferences with $k\!\ge\!2$ \cite{bennett1960multidimensional,davis1972social}. The Kemeny ranking problem becomes polynomial-time solvable if the preferences are single-peaked, single-crossing or 1-Euclidean, because the majority relation between candidates is then transitive and gives rise to a consensus ranking for the Kemeny rule. 

As the Kemeny ranking problem becomes easy for 1-Euclidean preferences, a natural subsequent question is whether this result also holds for $k$-Euclidean preferences with $k\!\ge\!2$. In~\cite{Thekla2021ADT} a restricted version of the problem, where the consensus (Kemeny) ranking is required to be also embeddable in the same euclidean plan as the preferences of the profile, is shown to be polynomial time solvable. In this note, we deal with the general problem and show that Kemeny ranking is NP-hard for 2-Euclidean preferences (and that the decision version is NP-complete), hence for $k$-Euclidean preferences with any $k\!\geq\!2$. Moreover, we show that it is still the case under norms $\ell_1$ and  $\ell_\infty$, i.e., if we consider preferences that can be represented in a 2-dimensional space using these norms (see Section~\ref{sec:def} for a formal definition). This latter setting, which is a variant of the usual $k$-Euclidean preferences, has been advocated in particular by Eguia~\cite{eguia2011foundations}, and Peters~\cite{peters2017recognising} recently showed that the problem of recognising preference profiles that are $d$-Euclidean with respect to $\ell_1$ or $\ell_\infty$ is in NP for $d\!\ge\!1$, leaving as an open question whether the problem is polynomial-time solvable or not.

In the present note we prove the following result:
\begin{theorem}\label{th1}
Under norms  $\ell_1$, $\ell_2$ and $\ell_\infty$, Kemeny ranking on 2-dimensional Euclidean preferences is NP-hard. This is true even if a 2-dimensional representation of preferences is given in the input.
\end{theorem}

We actually prove Theorem~\ref{th1} by showing a version of Debord's theorem \cite{debord1987caracterisation} (refining McGarvey's theorem \cite{mcgarvey1953theorem}) for Euclidean preferences (see Section~\ref{subsec:tournament} for a precise definition), namely:

\begin{theorem}\label{lemma:mcgarvey}
Every weighted tournament with weights of the same parity is inducible by a 2-dimensional $\ell_2$-Euclidean profile. Every weighted bipartite tournament with even weights is inducible by  a 2-dimensional $\ell_1$-Euclidean profile, and by a 2-dimensional  $\ell_\infty$-Euclidean profile.
\end{theorem}

Thus, essentially, hardness results for computational social choice problems that can be formulated on the (weighted) majority tournament are still true if preferences are 2-dimensional Euclidean because this assumption is not restrictive with regard to the weighted majority tournament. 
In particular, this is the case for the  {\it Slater rule}, which asks for  a consensus ranking which minimizes the number of disagreements with pairwise majority comparisons \cite{slater1961inconsistencies}. While this rule is often considered as a tournament solution concept (the Slater set consists of the winning candidates), it also defines a consensus ranking given a preference profile \cite{conitzer2006computing}. Using Theorem~\ref{lemma:mcgarvey}, we get the following result. %While this rule is often considered as a tournament solution (the Slater set consists of the winning candidates), it also defines a consensus ranking given a preference profile \cite{conitzer2006computing}. We note that,
%as the (positive) majority margins are all equal to 2 in the preference profiles generated in the reductions that will be used to prove Theorem~\ref{th1}, %and thus 
%the Kemeny rule and the Slater rule coincide for these profiles \cite{fischer2016weighted}.  Thus, we have the following Corollary.

\begin{corollary}
Under norms $\ell_1$, $\ell_2$ and $\ell_\infty$, Slater ranking on 2-dimensional Euclidean preferences is NP-hard. This is true even if a 2-dimensional representation of preferences is given in the input.
\end{corollary}

For the sake of brevity, we only mention the Kemeny rule in the remainder of the paper.

The note is organized as follows. Section~\ref{sec:def} is dedicated to preliminary definitions and notations. Section~\ref{sec:l1}, \ref{sec:linf} and \ref{sec:l2} are devoted respectively to the presentation of the proof of Theorem~\ref{lemma:mcgarvey} under norms $\ell_1$, $\ell_\infty$ and $\ell_2$.

\section{Definitions and notations}\label{sec:def}

Before establishing our main results, we formally define the Kemeny ranking problem, as well as $k$-Euclidean preferences under norms $\ell_1$, $\ell_2$ and $\ell_\infty$, and the notions related to tournaments.

\subsection{Kemeny ranking}

We consider a set $C$ of candidates and a set $V$ of voters. Each voter $v$ in $V$ ranks all the candidates (total rankings, no ties). The ranking of voter $v$ is denoted by $>_v$, where we write $c_i >_v c_j$ if $v$ prefers $c_i$ to $c_j$. The set of preferences of voters on candidates is called a (preference) profile.\\

Given two rankings $>_1$ and $>_2$ on candidates, let $kt(>_1,>_2)$ denote the Kendall tau distance between $>_1$ and $>_2$, i.e., the number of pairs of candidates $\{c_i,c_j\}$ such that $c_i >_1 c_j$ and $c_j >_2 c_i$, or vice versa ($c_i >_2 c_j$ and $c_j >_1 c_i$). The Kendall tau distance $KT(>,\mathcal{P})$ between a ranking $>$ and a profile $\mathcal{P}$ is then defined as:
\vspace{-0.10cm}
$$KT(>,\mathcal{P})=\sum_{>_v\in \mathcal{P}} kt(>,>_v)$$

Now, we are able to define the Kemeny ranking problem:

\begin{definition}[Kemeny ranking]
In the Kemeny ranking problem, given a preference profile $\mathcal{P}$, we want to determine a ranking $>$ on the candidates that minimizes $KT(>,\mathcal{P})$.
\end{definition}

In the decision version of Kemeny ranking,  given some integer $k$, we want to determine whether there exists a ranking $>$ such that $KT(>,\mathcal{P})\!\leq\!k$, or not. As stated in the introduction, we recall that this problem is known to be NP-complete \cite{bartholdi1989voting}.

\subsection{Euclidean preferences under norms $\ell_1$, $\ell_2$ and $\ell_\infty$}

Let $k\geq 1$ be an integer. We recall that given two points $p,q\in \mathbb{R}^{k}$:
\begin{itemize}
    \setlength\itemsep{-0.5em}
    \item $\ell_1$ is the norm associated to the distance $d_1(p,q)=\sum_{j=1}^{k} |p_j-q_j|$, where $p_j$ is the value of $p$ on the $j$-th coordinate;
    \item $\ell_2$ is the norm associated to the distance $d_2(p,q)=\sqrt{\sum_{j=1}^{k} (p_j-q_j)^2}$;
    \item $\ell_\infty$ is the norm associated to the distance $d_\infty(p,q)=\max_{j=1}^{k} |p_j-q_j|$.
\end{itemize}

We can now define $k$-Euclidean preferences under  norm $\ell_p$: 

\begin{definition}
Let $k\!\geq\!1$ be an integer, and $p\!\in\!\{1,2,\infty\}$. A profile $\mathcal{P}$ over sets $C$ of candidates and $V$ of voters  is $k$-Euclidean under  norm $\ell_p$ if there exists a mapping 
$h\!:\!C\!\cup\!V\!\rightarrow\!\mathbb{R}^{k}$ such that for each $v\!\in\!V$ and each $\{c_i, c_j\}\!\subseteq\!C$ ($i\!\neq\!j$): 
$$ c_i >_v c_j \Longleftrightarrow d_p(h(v),h(c_i))< d_p(h(v),h(c_j))$$
\end{definition}

Given the positions of voters and candidates in $\mathbb{R}^{k}$, the norm considered obviously has a very strong influence on the preferences, as illustrated by the following example.

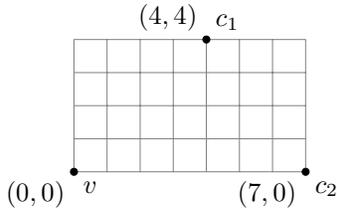
\begin{figure}[ht]
\begin{center}
\begin{tikzpicture}[scale=0.44]
\draw [very thin, gray] (0,0) grid (7,4);

\filldraw[black] (0,0) circle (3pt) node[anchor=north west] {$v$}
node[anchor=north east] {$(0,0)$};

\filldraw[black] (4,4) circle (3pt) node[anchor=south west] {$c_1$}
node[anchor=south east] {$(4,4)$};

\filldraw[black] (7,0) circle (3pt) node[anchor=north west] {$c_2$}
node[anchor=north east] {$(7,0)$};
\end{tikzpicture}
\caption{A voter $v$ and two candidates $c_1,c_2$ in $\mathbb{R}^2$.}
  \label{fig:exEuclidean}
\end{center}
\end{figure}

\begin{example}
Consider voter $v$ and two candidates $c_1$ and $c_2$, with $h(v)\!=\!(0,0)$, $h(c_1)\!=\!(4,4)$ and $h(c_2)\!=\!(7,0)$, as illustrated in Figure~\ref{fig:exEuclidean}. Under norm $\ell_2$, the preference $c_1 >_v c_2$ holds, while, on the contrary, $c_2 >_v c_1$ under norm $\ell_1$.
\end{example}

Note that $1$-Euclidean preferences under norms $\ell_1$, $\ell_2$ and $\ell_\infty$ are equivalent, as $d_1(p,q)\!=\!d_2(p,q)\!=\!d_\infty(p,q)$ if $k\!=\!1$. Furthermore, $1$-Euclidean preferences are both single-peaked and single-crossing \cite{grandmont1978intermediate}, but do not coincide with the set of single-peaked single-crossing preferences\footnote{A profile is \emph{single-peaked} if there exists an axis on the candidates such that the preferences of each voter are decreasing as one moves away from his/her most-preferred candidate along the axis, and \emph{single-crossing} if there exists an axis on the voters such that, for each couple $(c_i,c_j)$ of candidates, the set of voters $v$ for which $c_i\!>_v\!c_j$ is connected along the axis.}, as there are examples of such profiles that are not 1-Euclidean \cite{elkind2014characterization}. While there are polynomial-time algorithms for recognizing 1-Euclidean preferences \cite{doignon1994polynomial,knoblauch2010recognizing}, recognizing $k$-Euclidean preferences under norm $\ell_2$ is NP-hard for $k\!\ge\!2$ \cite{peters2017recognising}. Consequently, in the rest of the note, we assume that the positions of the candidates and the voters in $\mathbb{R}^{k}$ are given as part of the input in the Kemeny ranking problem.

\subsection{Weighted tournaments}\label{subsec:tournament}

A tournament is a directed graph $G=(N,A)$ where for each pair $\{u_i,u_j\}$ of vertices there is exactly one arc (either $(u_i,u_j)$ or $(u_j,u_i)$). The name tournament refers to a situation where a game (without tie) is organised between each pair of nodes $\{u_i,u_j\}$, and the arc represents who won the game (say $(u_i,u_j)$ if $u_i$ wins). Weighted tournaments are then a generalization where each game is won by a certain margin, which defines the integer weight $w(u_i,u_j)\!>\!0$ of arc $(u_i,u_j)$ (if $u_i$ wins). Note that, if $u_i$ and $u_j$ are ex \ae{}quo, then $w(u_i,u_j)\!=\!0$ and there is no arc between $u_i$ and $u_j$. A weighted tournament is bipartite if the corresponding digraph is bipartite.

\begin{definition}
A weighted tournament on a set $N=\{u_1,\dots,u_n\}$ of nodes is {\it inducible} if there exists a preference profile $\mathcal{P}=(C,V)$ on a set $C=\{c_1,\dots,c_n\}$ of candidates such that for any pair $\{u_i,u_j\}$ of nodes, $w(u_i,u_j)=|\{v\in V: c_i>_v c_j\}| - |\{v\in V: c_j>_v c_i\}|$.
\end{definition}

It is known \cite{mcgarvey1953theorem,debord1987caracterisation} that a weighted tournament is inducible if and only if all the weights are of the same parity. In the following, we refer to odd (resp. even) weighted tournaments if all the weights are odd (resp. even). Note that when $|N|\geq 3$ a bipartite weighted tournament is necessarily even (as there is at least one pair $\{u_i,u_j\}$ with $w(u_i,u_j)=0$).

As mentioned in the introduction, it is also well known that showing that any tournament (or bipartite tournament) is inducible (using a polynomial time construction) allows to derive NP-hardness for Kemeny ranking, using a reduction from the feedback arc set problem - the decision version of which is NP-complete also in bipartite graphs.

A classical way to build a profile that induces a given weighted tournament result is to convert the nodes of $G$ into candidates and the arcs into voters. More precisely, consider an even\footnote{If the tournament is odd, then we can add one voter to the construction, and use pairs $f_{ij}$ and $g_{ij}$ to adjust the weights as needed.} weighted tournament, and suppose that we build a profile  such that:
\begin{enumerate}
    \item There is one candidate $c_i$ for each vertex $u_i$;
    \item  For each arc $(u_i,u_j)$, there are $w(i,j)/2$ (identical) copies of a voter $f_{ij}$ and $w(i,j)/2$ (identical) copies of a voter $g_{ij}$;
    \item (All copies of) both voters $f_{ij}$ and $g_{ij}$ prefer $c_i$ to $c_j$;
    \item For any other pair $\{c,d\}$ of candidates, among the two voters $f_{ij}$ and  $g_{ij}$, exactly one   prefers $c$ to $d$ (and the other one prefers $d$ to $c$).
\end{enumerate}
Then such a profile clearly induces the desired weighted tournament.

If preferences are unrestricted, such properties for the preferences of $f_{ij}$ and $g_{ij}$ can be obtained, for instance, by following the approach proposed by McGarvey \cite{mcgarvey1953theorem}, consisting in defining the preferences of $f_{ij}$ as $c_i\!>\!c_j\!>\!c_1\!>\!c_2\!>\!\ldots\!>\!c_n$ and the preferences of $g_{ij}$ as $c_n\!>\!c_{n-1}\!>\!\ldots\!>\!c_1\!>\!c_i\!>\!c_j$. In the sequel, we show that we can still obtain the previous properties with an Euclidean profile under norms $\ell_1$, $\ell_2$ or $\ell_\infty$.

\section{Proof of Theorem~\ref{lemma:mcgarvey} under $\ell_1$}
\label{sec:l1}

We start with a weighted bipartite tournament (thus necessarily even, as mentioned earlier), where $G$ is a bipartite graph with vertex set $L\cup R$ and arc set $A$ (each arc having one extremity in $L$ and one in $R$). We denote by $n$ the number of vertices, and by $m$ the number of arcs.

\begin{figure}[ht]
\begin{center}
\begin{tikzpicture}[scale=0.44]

\draw [-] (2,2) -- (10,2);
\draw [-] (2,10) -- (10,10);
\draw [-] (2,2) -- (2,10);
\draw [-] (10,2) -- (10,10);

\draw [-] (2.5,4.2) -- (2.5,9.5);
\draw [-] (2.5,9.5) -- (4.8,9.5);
\draw [-] (9.5,6.2) -- (9.5,9.5);
\draw [-] (9.5,9.5) -- (5.2,9.5);

\draw [-] (2.5,3.8) -- (2.5,2.5);
\draw [-] (2.5,2.5) -- (6.8,2.5);
\draw [-] (9.5,5.8) -- (9.5,2.5);
\draw [-] (9.5,2.5) -- (7.2,2.5);

\filldraw[black] (2,4) circle (3pt) node[anchor=east] {$c_i$};

\filldraw[black] (10,6) circle (3pt) node[anchor=west] {$c_j$};

\filldraw[black] (5,10) circle (3pt) node[anchor=south] {$A_{ij}$};

\filldraw[black] (7,2) circle (3pt) node[anchor=north] {$B_{ij}$};

\filldraw[black] (4.05,10) circle (3pt) node[anchor=south] {$f_{ij}$};

\filldraw[black] (6.15,2) circle (3pt) node[anchor=north] {$g_{ij}$};

\end{tikzpicture}
\caption{The construction with two vertices $u_i,u_j$ and an arc $(u_i,u_j)$.}
  \label{fig:rectangle}
\end{center}
\end{figure}
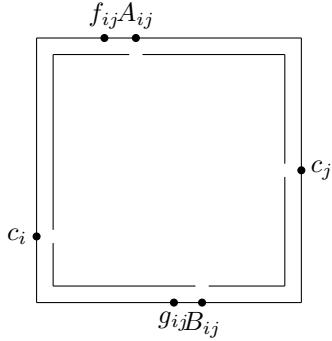

We build an instance where candidates and voters lie on a square, whose sides are parallel to the axes 
(see Figure~\ref{fig:rectangle}). More precisely:
\begin{itemize}
    \item Each vertex $u_i$ corresponds to a candidate $c_i$. If $u_i\!\in\!L$ (resp. $u_i\!\in\!R$), $c_i$ will be on the vertical left side (resp. right side) of the square. We will say that $c_i\!\in\!L$ (resp. $c_i\!\in\!R$) if $u_i\!\in\!L$ (resp. $u_i\!\in\!R$)
    \item Each arc $(u_i,u_j)$ corresponds to $w(i,j)/2$ (identical) copies of a voter $f_{ij}$ and $w(i,j)/2$ identical copies of a voter $g_{ij}$. Point $f_{ij}$ will be on the horizontal upper side of the square, while $g_{ij}$ will be on the horizontal lower side of the square. 
\end{itemize}

Let us consider an arc $(u_i,u_j)$, with $u_i\!\in\!L$ and $u_j\!\in\!R$. We call $A_{ij}$ the point on the upper horizontal side such that $d_1(c_i,A_{ij})\!=\!d_1(c_j,A_{ij})$ (note that such a point indeed exists on the upper horizontal side of the square). Similarly, we call $B_{ij}$ the point on the lower horizontal side such that $d_1(c_i,B_{ij})=d_1(c_j,B_{ij})$.

We put the voters $f_{ij}$  (resp. $g_{ij}$)  on the upper (resp. lower) horizontal side at $\epsilon$ (to be specified) to the left of $A_{ij}$ (resp. of $B_{ij}$). If the arc had been $(u_j,u_i)$, then the voters $f_{ji}$ and $g_{ji}$ would have been at $\epsilon$ to the right of $A_{ij}$ and $B_{ij}$.

Assume that we choose the vertical positions of candidates in such a way that all $A_{ij}$ are distinct (and equivalently, all $B_{ij}$ are distinct) - we give later an explicit construction that fulfills this condition. Note that as $B_{ij}$ and $A_{ij}$ are symmetric with respect to the center of the square, the order of $A$-points on the upper side is the inverse order of $B$-points on the lower side. 

Then we choose $\epsilon$ sufficiently small so that between $f_{ij}$ and $A_{ij}$ there is no other $A$-point (for instance, if $x_{ij}$ denotes the $x$-coordinate of $A_{ij}$, we can choose $\epsilon=\frac{\min\{|x_{ij}-x_{kl}|\}}{3}$), where the minimum is taken over all pairs of (distinct) points $A_{ij},A_{k\ell}$). Similarly, between $g_{ij}$ and $B_{ij}$ there is no other $B$-point.

Let us consider an arc $(u_i,u_j)$ with $u_i\in L$ and $u_j\in R$ (the other case being completely symmetric). Then:
\begin{itemize}
    \item Both voters $f_{ij}$ and $g_{ij}$ prefer $c_i$ to $c_j$ (as $A_{ij}$ is equidistant from $c_i$ and $c_j$, the same for $B_{ij}$).
    \item For any other pair $\{c,d\}$ of candidates, exactly one voter among $f_{ij}$ and $g_{ij}$ prefers $c$ to $d$ (and one prefers $d$ to $c$). This is easy to see if both $c$ and $d$ belong to $L$, or if both belong to $R$. If $c\!\in\!L$ and $d\!\in\!R$, if for instance their corresponding $A$-point is on the right of $A_{ij}$, then their corresponding $B$-point is on the left of $B_{ij}$, thus $f_{ij}$ prefers $c$ to $d$ but $g_{ij}$ prefers $d$ to $c$. 
\end{itemize}

Thus, this construction fulfills the conditions 1--4 given in Subsection~\ref{subsec:tournament}, and yields a profile inducing the desired (bipartite) tournament. We now give an explicit (polynomial time) construction which ensures that $A$-points are distinct, as well as $B$-points.\\

{\bf Explicit construction}

We consider a square with  side lengths $\Delta \!=\!2^{n+1}$, where $n$ is the number of vertices of the graph. Let us consider that the bottom left corner of the rectangle has coordinates $(0,0)$.

We set the $y$-coordinate of candidate $c_i$ to $y_i\!=\!2^i$. Then the $x$-coordinate $x_{ij}$ of $A_{ij}$ is such that $x_{ij}+\Delta  - y_i=\Delta-x_{ij}+\Delta -y_j$, meaning that:
$$x_{ij}=\frac{\Delta +y_i-y_j}{2}.$$
Then we can verify that these $y$-values are such that all $A$-points are distinct. Indeed, for any distinct pairs $\{i,j\}$ and $\{k,\ell\}$ of indices, $y_i-y_j\neq y_k-y_\ell$, i.e., $y_i+y_\ell\neq y_k+y_j$. To see this, if say $\ell$ is the largest among the indices, then:
\begin{itemize}
    \item If $j=\ell$, then $i\neq k$ (as the pairs are distinct), and $y_i+y_\ell\neq y_k+y_j$. 
    \item If $j<\ell$, then as $k<\ell$ we have $y_k+y_j\leq 2(2^{\ell-1})=2^\ell=y_\ell<y_i+y_\ell$.
\end{itemize}

Then all the values $x_{ij}=\frac{\Delta+y_i-y_j}{2}$ are distinct. Note that as $y$-values and $\Delta$ are even integers, $x_{ij}$ is an integer, and we can choose $\epsilon\!=\!\frac{1}{2}$ (and multiply everything by 2 if we want integers).

As the coordinates can be encoded with a polynomial number of bits, the reduction is polynomial time. 

\section{Proof of Theorem~\ref{lemma:mcgarvey} under $\ell_\infty$}
\label{sec:linf}

\begin{figure}[ht]
\begin{center}
\begin{tikzpicture}[scale=0.3]
\draw [->] (-10,0) -- (10,0);
\draw [->] (0,-10) -- (0,10);
\draw [-] (8,0) -- (0,8);
\draw [-] (0,8) -- (-8,0);
\draw [-] (-8,0) -- (0,-8);
\draw [-] (0,-8) -- (8,0);
\filldraw[black] (8,0) circle (3pt) node[anchor=north] {$\Delta$};
\filldraw[black] (0,-1) circle (0pt) node[anchor=east] {O};
\filldraw[black] (2,6) circle (3pt) node[anchor=west] {$c_j$};
\filldraw[black] (0,8) circle (3pt) node[anchor=east] {$\Delta$};
\filldraw[black] (-5,-3) circle (3pt) node[anchor=east] {$c_i$};
\filldraw[black] (-4.5,3.5) circle (3pt) node[anchor=south] {$A_{ij}$};
\filldraw[black] (4.5,-3.5) circle (3pt) node[anchor=south] {$B_{ij}$};
\end{tikzpicture}
\caption{The construction with two vertices $v_i$, $v_j$.}\label{fig:carrepenche}
\end{center}
\end{figure}
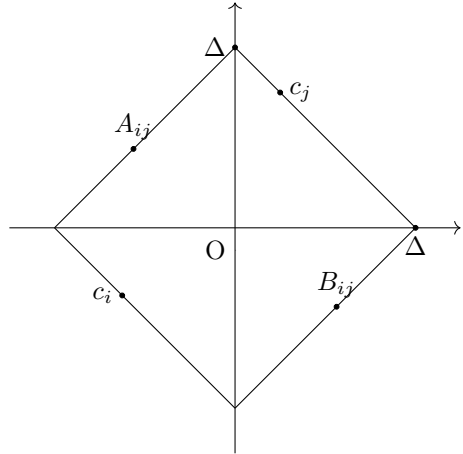

We use a construction which is similar to the case of $\ell_1$, but positioning candidates and voters on a square which is oriented as in Figure~\ref{fig:carrepenche}. The diagonal of the square has length $2\Delta$ with $\Delta\!=\!2^{n+1}$.

We position a candidate $c_i\!\in\!L$ on the lower left side, at position $(-2^i,2^i-\Delta)$. 
A candidate $c_j\!\in\!R$ is on the upper right side, at position $(2^j,\Delta-2^j)$.

Then we define two points $A_{ij}$ and $B_{ij}$, respectively on the upper left side and on the lower right side, both being equidistant (under $\ell_\infty$) from $c_i$ and $c_j$. Namely, the coordinates of $A_{ij}$ are $(\frac{2^i+2^j}{2}-\Delta,\frac{2^i+2^j}{2})$.  
Point $B_{ij}$ is such that $B_{ij}$ and $A_{ij}$ are symmetric with respect to the center O of the square. 

As previously, if there is an arc $(u_i,u_j)$ with $u_i\!\in\!L$ and $u_j\!\in\!R$, we create $w(u_i,u_j)/2$ identical copies of two voters $f_{ij}$ and $g_{ij}$, point $f_{ij}$ being positioned on the upper left side at $\epsilon$ to the bottom/left of $A_{ij}$, and $g_{ij}$ being positioned on the lower right side  at $\epsilon$ to the bottom/left of $B_{ij}$. If there is an arc $(u_j,u_i)$ with $u_i\!\in\!L$ and $u_j\!\in\!R$, then $f_{ji}$ and $g_{ji}$ are positioned at $\epsilon$ to the right/up of $A_{ij}$ and $B_{ij}$ respectively.

The choice of the coordinates of candidates ensures that all the $A$-points and $B$-points are distinct (for the same reason as in the proof for the $\ell_1$ norm), and integral. Then we choose $\epsilon$ sufficiently small ($\epsilon\leq 1/2$) so that between $f_{ij}$ and $A_{ij}$ there is no other $A$-point. Similarly, between $g_{ij}$ and $B_{ij}$ there is no other $B$-point. %so we can choose $\epsilon\!=\!1/2$ to ensure that there \textcolor{blue}{are} no $A$-points between $f_{ij}$ and $A_{ij}$ (neither $B$-points between $g_{ij}$ and $B_{ij}$).

\section{Proof of Theorem~\ref{lemma:mcgarvey} under $\ell_2$}
\label{sec:l2}

\begin{figure}[ht]
\begin{center}
\begin{tikzpicture}[scale=0.2]

\draw (0,0) circle (10) ;
\draw[-] (-10,-10) -- (10,10);

\filldraw[black] (3.122,9.5) circle (3pt) node[anchor=south] {$c_i$};

\filldraw[black] (9.5,3.122) circle (3pt) node[anchor=west] {$c_j$};

\filldraw[black] (7.071,7.071) circle (3pt) node[anchor=west] {$A_{ij}$};

\filldraw[black] (-7.071,-7.071) circle (3pt) node[anchor=north] {$B_{ij}$};

\filldraw[black] (6.5,7.599) circle (3pt) node[anchor=south] {$f_{ij}$};

\filldraw[black] (-7.599,-6.5) circle (3pt) node[anchor=east] {$g_{ij}$};

\filldraw[black] (0,0) circle (3pt) node[anchor=north] {O};

\filldraw[black] (4.5,4.5) circle (0pt) node[anchor=east] {$D_{ij}$};

\end{tikzpicture}
\caption{The construction with two vertices $v_i,v_j$ and an arc $(v_i,v_j)$}
  \label{fig:circle}
\end{center}
\end{figure}
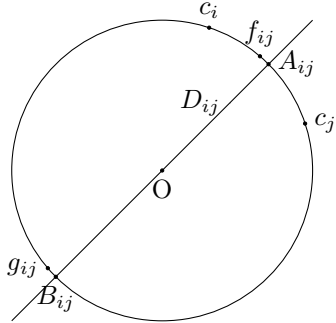

 We now start with an even weighted tournament, and will position candidates and voters on a circle, centered at point O of coordinates $(0,0)$. More precisely (see Figure~\ref{fig:circle}):
\begin{itemize}
    \item Each vertex $u_i$ corresponds to a candidate $c_i$ positioned on the circle.
    \item Let us call $D_{ij}$ the  line of equidistant points (under $\ell_2$) between $c_i$ and $c_j$, and $A_{ij}$ and $B_{ij}$ the two points of $D_{ij}$ on the circle. Each arc $(u_i,u_j)$ corresponds to $w(i,j)/2$ identical copies of a voter $f_{ij}$ and $w(i,j)/2$ identical copies of a voter $g_{ij}$, all positioned on the circle. Point $f_{ij}$ is on the same side of $D_{ij}$ as $c_i$, with vectors $\overrightarrow{OA_{ij}}$ and $\overrightarrow{Of_{ij}}$ forming an angle of $\epsilon$ (in absolute value). Similarly,  $g_{ij}$ is on the same side of $D_{ij}$ as $c_i$, with vectors $\overrightarrow{OB_{ij}}$ and $\overrightarrow{Og_{ij}}$ forming an angle of $\epsilon$ (in absolute value).
\end{itemize}

As previously, suppose that we choose the positions of candidates in such a way that all the points $A_{ij}$ and $B_{ij}$ are distinct. 

Then we choose $\epsilon$ sufficiently small such that between $f_{ij}$ and $A_{ij}$ there is no $A$-point or $B$-point, and similarly between $g_{ij}$ and $B_{ij}$ there is no $A$-point or $B$-point.

Let us consider an arc $(u_i,u_j)$. Then:
\begin{itemize}
    \item Both voters $f_{ij}$ and $g_{ij}$ prefer $c_i$ to $c_j$ (as $A_{ij}$ is equidistant from $c_i$ and $c_j$, the same for $B_{ij}$).
    \item For any other pair $\{c,d\}$ of candidates, exactly one voter among $f_{ij}$ and $g_{ij}$ prefers $c$ to $d$ (and one prefers $d$ to $c$). This follows from the fact that all $D$-lines intersect in $O$, meaning that $f_{ij}$ and $g_{ij}$ cannot be on the same side of the $D$-line corresponding to $\{c,d\}$.
\end{itemize}

{\bf Explicit construction}

Let us call $\Theta_i$ the angle (polar coordinate, in radian) of $c_i$ (i.e., the angle between the horizontal axis and $\overrightarrow{Oc_i}$). Then we shall choose $\Theta_i$ in such a way that all the points $A_{ij}$ and $B_{ij}$ are distinct. This appears as soon as $(\Theta_i+\Theta_j)$ are distinct for all choices of distinct pairs $\{c_i,c_j\}$, as the angle of the line $D_{ij}$ is $\frac{\Theta_i+\Theta_j}{2}$. 

Let us fix $\Theta_i=\frac{2^i}{2^n}=2^{i-n}$. By the same reasoning as in the proof for the $\ell_1$ norm, all $(\Theta_i+\Theta_j)$ are distinct (note that $0\leq \Theta_i\leq \pi/2$ so  $(\Theta_i+\Theta_j)$ are indeed distinct modulo $2\pi$). We can fix $\epsilon=1/2^{n+1}$, to fulfill the property for $f_{ij}$ and $g_{ij}$.

We note that the actual preference profile can be easily built from this embedding of points in the 2-dimensional space. Indeed, if $i>j$, voter $f_{ij}$ has angle $\frac{2^i+2^j+1/2}{2^n}$, and she prefers $c_k$ to $c_l$ iff $|2^k-a_{ij}|<|2^l-a_{ij}|$, where $a_{ij}=2^i+2^j+1/2$ (if $i<j$ it is the same with $a_{ij}=2^i+2^j-1/2$). Voter $g_{ij}$ has the reverse preference on all pairs but $\{c_i,c_j\}$. Thus, the reduction is polynomial time.

\section{Conclusion}

In this note, we have proved that the result of McGarvey~\cite{mcgarvey1953theorem} and Debord~\cite{debord1987caracterisation} about inducible weighted tournaments is still true for Euclidean preferences under norm $\ell_2$, and that every even weighted bipartite tournament is inducible by Euclidean preferences under norm $\ell_1$ and $\ell_\infty$. These results allowed us to answer an open question, namely that regarding the computational complexity of the Kemeny ranking problem when the input preferences are $k$-Euclidean with $k\!\ge\!2$. We have proved that the problem remains NP-hard, contrary to the case of 1-Euclidean preferences, under norms $\ell_1$, $\ell_2$ and $\ell_\infty$. Another consequence of our results is that computing an optimal consensus ranking for the Slater rule is also NP-hard for $k$-Euclidean preferences with $k\!\ge\!2$ under the same norms.

Natural research directions to pursue would be:
\begin{itemize}
    \item to generalize the result about Kemeny ranking to other Minkowski $\ell_p$ norms. We conjecture that the problem is NP-hard as well for $p\neq 1,2,\infty$.
    \item to investigate the impact of 2-dimensional Euclidean preferences on the complexity of other NP-hard social choice problems that cannot be formulated on an induced weighted tournament; for instance, Godziszewski et al. \cite{godziszewski2021analysis} showed that computing the result of a number of multiwinner voting rules remains NP-hard with 2-dimensional Euclidean preferences, without resorting to weighted majority tournaments.
\end{itemize}

\end{document}